# *SoundSignature*: What Type of Music Do You Like?

**Brandon James Carone**
Department of Psychology, New York University
Music and Audio Research Laboratory (MARL), NYU
Center for Language Music and Emotion (CLaME), NYU
New York, New York
bcarone@nyu.edu

**Pablo Ripollés**
Department of Psychology, New York University
Music and Audio Research Laboratory (MARL), NYU
Center for Language Music and Emotion (CLaME), NYU
New York, New York
pripolles@nyu.edu

*Abstract*—SoundSignature is a music application that integrates a custom OpenAI Assistant to analyze users' favorite songs. The system incorporates state-of-the-art Music Information Retrieval (MIR) Python packages to combine extracted acoustic/musical features with the assistant's extensive knowledge of the artists and bands. Capitalizing on this combined knowledge, *SoundSignature* leverages semantic audio and principles from the emerging Internet of Sounds (IoS) ecosystem, integrating MIR with AI to provide users with personalized insights into the acoustic properties of their music, akin to a musical preference personality report. Users can then interact with the chatbot to explore deeper inquiries about the acoustic analyses performed and how they relate to their musical taste. This interactivity transforms the application, acting not only as an informative resource about familiar and/or favorite songs, but also as an educational platform that enables users to deepen their understanding of musical features, music theory, acoustic properties commonly used in signal processing, and the artists behind the music. Beyond general usability, the application also incorporates several well-established open-source musician-specific tools, such as a chord recognition algorithm (CREMA), a source separation algorithm (DEMUCS), and an audio-to-MIDI converter (basic-pitch). These features allow users without coding skills to access advanced, open-source music processing algorithms simply by interacting with the chatbot (e.g., can you give me the stems of this song?). In this paper, we highlight the application's innovative features and educational potential, and present findings from a pilot user study that evaluates its efficacy and usability.

## I. Introduction

The rise of digital media platforms in the 21st century has made music more accessible than ever, dramatically transforming how we consume and engage with it [1]. Globally, people listen to over 18 hours of music per week, facilitated by music streaming services and social media platforms that allow music lovers to explore, share, and enjoy millions of songs at their fingertips [2-4]. This shift from physical to digital media has not only profoundly affected the way we listen to music [1, 5, 6], but also the type of music we listen to: our daily listening habits are now heavily influenced by recommendation algorithms developed by streaming services [7-10]. While music streaming platforms collect a diverse plethora of data to create compelling and individualized experiences, users are often unaware of the reasons behind the music suggestions made by a recommender system [11, 12]. To inform users about their musical behaviors, products such as Spotify Wrapped and Apple Music Replay, for example, focus on painting a picture of each user's listening history throughout the year. However, they fail to dive deeper into what encourages users to return to the songs and artists they love most and the reasons why individuals use different types of music in their daily lives [13, 14].

Over the last two decades, advancements in Music Information Retrieval (MIR) techniques have opened several new avenues for computer programmers and computational musicologists to decompose, analyze, and alter music and audio signals [15]. One of the problems with these existing tools is that they require a degree of technical expertise that can become an entry barrier for those who lack a background in signal processing and/or in coding skills. While remarkable efforts have been made to use music as a contextual tool to facilitate the learning of the technical foundations required to understand signal processing (e.g., capitalizing on music to teach linear algebra, Fourier analysis, etc.) [16], many times these interactive lessons still require coding skills. This barrier contributes to the significant gap in the user's understanding of the acoustic characteristics of the music they love, as many music listeners might be eager to explore the deeper aspects of their favorite songs but lack the skills to do so effectively. Moreover, in spite of recent efforts in promoting public musicology [17] and musical analysis, both traditional and computational musicology [18] remain confined to academic or professional settings, limiting accessibility for general users. Thus, there remains a critical need for resources that are both easy to use and computationally efficient, removing the technical barriers while at the same time providing advanced music-related insights.

This paper introduces *SoundSignature*, a music analysis web-application that leverages state-of-the-art MIR and Natural Language Processing (NLP) technologies. After a user uploads a sample of their favorite songs, the application extracts various musical and acoustic features from each musical excerpt and feeds the results to a customized OpenAI Assistant [19]. The chatbot then combines – in a way that is both accessible and user-friendly – the extracted features (e.g., BPM, Spectral Centroid, Spectral Flux, etc.) with its extensive knowledge of the artists behind the songs to provide insights into the user's musical preferences. This system not only identifies musical elements but contextualizes them, enabling users to explore their musical tastes on a nuanced level, aligning with current research that emphasizes personalization and context-awareness in music recommendation systems [20]. *SoundSignature* focuses on providing users with a personalized overview of the specific acoustic and musical characteristics that make up their favorite songs, akin to a musical preference personality report. We believe this feature sets *SoundSignature* apart from other products that simply provide a summary of the listening preferences of the user (e.g., Spotify Wrapped). This emphasis on personalization aims to provide users with the feeling of being catered to and understood, thus enhancing their music experience and providing a description of their musical taste that goes beyond stating which musical genres or artists they like or have listened to the most. Another unique characteristic of *SoundSignature* is that it provides a brief description of the lyrical content and cultural context of each song and artist, respectively. Lyrics are often overlooked in music recommendation systems and personalized products like Spotify Wrapped. Therefore, incorporating OpenAI's vast knowledge of song lyrics and information about the artists might help users to better understand the meaning or emotion behind their favorite music [20, 21].

In addition to offering insights, *SoundSignature* encourages users to delve deeper into the analyses performed by allowing them to ask the chatbot further questions about the songs or the analyses described. Much like Mueller's approach to simplifying signal processing with musical coding examples [16], here we seek to further democratize signal processing by not only using music as a contextual tool to catch the users' attention and motivate them, but also by employing the users' favorite songs to enhance understanding of complex acoustic and/or musical concepts like spectrograms, spectral flux, or pulse clarity, among others.

Finally, we implement various state-of-the-art open-source tools such as music source separation and chord recognition algorithms, empowering users without coding skills to engage directly with advanced music processing algorithms (e.g., a user can ask the app to extract the stems of a song). Cloud computing, with its ability to offload processing and provide real-time, scalable access to complex systems, offers a promising solution to make sophisticated music analysis accessible to a broader audience. This aligns with the ecosystem provided by the *Internet of Sounds* (IoS) and the *Internet of Musical Things* (IoMusT) frameworks, where interconnected devices and cloud-based platforms enable seamless access to complex music processing and analysis tools [22-25]. The MIR analyses used by *SoundSignature* are performed via a *Streamlit* web application, which facilitates real-time interaction and cloud-based processing. Utilizing cloud infrastructure ensures scalability and accessibility, allowing users to interact with the system from any device with an internet connection [22, 26, 27]. By simplifying access to these tools, *SoundSignature* not only contributes to the growing IoS ecosystem, but also aims to facilitate creative exploration and experimentation for users with higher musical knowledge but limited coding skills.

This project is motivated by three aims: 1) to provide music enthusiasts, regardless of their musical knowledge or technical skills, with deep, personalized insights into their musical tastes; 2) to serve as an educational platform for users to learn about complex musical and acoustic concepts; and 3) to be a one-stop shop for musicians to use musical tools such as music source separation and chord recognition algorithms that are otherwise inaccessible to users with limited coding skills. To evaluate the efficacy and usability of *SoundSignature* we conducted a pilot user study. The methodology of this study, alongside detailed descriptions of the technology and algorithms employed, are explored in the following sections. This approach substantiates the application's capabilities and sets the stage for discussing its broader implications in reshaping how we interact with and understand music.

## II. SYSTEM ARCHITECTURE

The proposed system is composed of three main components: 1) a set of feature extractors, 2) tools for manipulating and processing the songs, and 3) an AI assistant that ties the tools together, synthesizing the extracted features with its knowledge of music to provide insights for the user. The entire system is presented to the user through a cloud-based web application. Thus, *SoundSignature*, situated within the broader context of the IoS and the IoMusT, interacts with musical data in a way that aligns with the IoS vision of connected, interactive, and intelligent sound devices. The following subsections describe the feature extraction pipeline, the additional tools integrated for musicians, the role of the AI assistant, and the web interface used to deliver the system.

### A. Musical / Acoustic Feature Extraction Pipeline

Feature extraction was performed with a variety of different music and signal processing packages, such as *librosa* 0.10.1 [28] and *essentia* 2.1.b6.dev117 [29] running in Python version 3.9. We began by using *librosa* to extract the audio signal and sampling rate, and then ran the short time Fourier transform on the audio signal. Following that, we output the spectrogram for the user to see. While many acoustic and/or musical features can be calculated, we capitalized on previous MIR, music psychology, and music neuroscience literature to select a few features representing core elements of music [30-32]. Specifically, we focused on features clustered across four categories: Tempo and Rhythmicity, Harmony and Melody, Timbre and Texture, and Energy and Dynamics. For Tempo and Rhythmic features, we focused on beats per minute (BPM; the speed or pace of the music) and pulse clarity (clarity of the rhythmic pulse or beat). For Harmonic and Melodic features, we extracted the key and mode (e.g., C major or A minor), as well as the key strength (clarity and stability of the detected musical key). For the Timbre and Texture features, we extracted the spectral centroid ("center of mass" of the spectrum, indicating brightness), spectral bandwidth (the width of significant frequency bands, indicating sound fullness), and spectral flux (the rate of change in the power spectrum, indicating musical onsets or texture changes). Finally, for the Energy and Dynamics features, we extracted RMS energy (average power of the audio signal independent of human perception) and loudness (psychoacoustic model that incorporates frequency weighting to reflect the human ear's varying sensitivity to different frequencies) [29]. The system's reliance on acoustic and musical feature extraction aligns with the interdisciplinary domain of semantic audio [25, 33], which uses signal processing and machine learning techniques to extract structured, meaningful information from music. By applying these techniques, *SoundSignature* provides users with high-level interpretations of their music preferences, facilitating human interaction with audio data through accessible, interpretable insights. This is aligned with the increasing demand for semantic audio features in the context of the IoS [22].

*B. Tools for Musicians*

In addition to the features extracted from each song, we have also included in the app some state-of-the-art music tools that can be especially useful for musicians and other individuals with a high musical knowledge (e.g., composers). We include tools for chord identification and stem and midi extraction. Specifically, the app includes *Meta's Hybrid Transformers for Music Source Separation (HT-DEMUCS) [34]*, the *Convolutional and Recurrent Estimators for Music Analysis (CREMA)* chord recognition algorithm [35], and *Spotify's Basic-Pitch audio-to-midi converter* [36]. Users can call the functions from these packages from the prompt by using specific words such as "stems," "chords," or "midi" (e.g., "can you give me the stems of this song?"; "create a midi file from this song."; "what are the chords in this song?"). After mentioning one of these keywords, a dropdown menu appears, and you select which of the MP3s you are referring to. Chords are output in a chart with "Start Time", "End Time", "Chord", and "Confidence" as the column headers, whereas the stems and midi files are both playable and downloadable from the web application (See Supplementary Materials for example).

*C. OpenAI Assistant*

In order to interpret the users' requests and the output of the extracted acoustic and musical features, we created a custom OpenAI Assistant running on the gpt-4o model by utilizing the OpenAI API [19]. Whereas the ChatGPT prompt is rather general, allowing it to adapt to any query a user might input, our customized assistant was informed that it is an advanced music analysis tool. Here is a condensed version of the prompt that we used:

"You are a sophisticated music analysis tool designed to offer personalized insights into users' musical preferences and what their favorite songs reveal about them. Users upload their favorite songs (format: SongName_ArtistName.mp3), and you analyze the similarities and differences using acoustic and musical features extracted with Python packages like *librosa* and *essentia*. Translate these technical features into relatable music characteristics, comparing and contrasting songs to highlight what defines the user's tastes. Emphasize patterns that indicate a preference for specific musical styles or elements and how these may reflect their personality. Provide insights into what users might seek in music, drawing from your knowledge of the artists and songs for a comprehensive analysis. Use varied descriptors to avoid repetition and adjust the complexity of the language based on user preferences, starting with simplified explanations. Conclude each analysis by asking if the user needs further clarification or more detailed technical explanations."

It is important to note that while all of the previously mentioned features are manually extracted from the MP3s in Python, the cultural and historical contexts and lyrical content are pulled from OpenAI's knowledge base, which was most recently updated in October 2023 [19].

*D. Building the Website*

To create the *SoundSignature* website and avoid translating the Python code to html, we utilized *Streamlit* [37], a Python package meant to host data science and machine learning applications. Users can edit some model parameters on the web application, enter their own custom OpenAI API Key, speak instead of type, or upload images, among others. The homepage instructs users to upload their songs in the specified format (SongName_ArtistName.mp3; see Figure 1A). After the user uploads their music, the analysis function is called, and the results are displayed for the user to view (see Figure 1B). Once the analyses have finished running, the extracted features are fed to the chatbot, whereby it then produces its response (see Figure 1CDE). When coding the website, we attempted to maintain a simple user interface while ensuring that things ran smoothly, much like the ChatGPT website. See also the Supplementary Materials for examples of full responses of the app. For those interested in using the *SoundSignature* application, please reach out via email to bcarone@nyu.edu to discuss access. We aim to accommodate users within the constraints of our current resources.

III. PILOT USER STUDY

To evaluate the efficacy and usability of *SoundSignature* we conducted a pilot user study in which we collected quantitative and qualitative responses from participants who used the app.

*A. Participants*

A total sample of 20 participants (7 females, mean age = 33.65, sd = 13.22) completed the study. For participant demographics, the sample consisted of 60% White, 25% Other, 5% East Asian, 5% Black or African American, and 5% who preferred not to answer. Regarding ethnicity, 70% were Hispanic or Latinx. In terms of education, 35% were enrolled in or had completed a graduate program, 45% held a Bachelor's degree, 10% had some college education, and 10% completed some of high school. Informed consent was collected from each participant, and the study was completed in accordance with the NYU Institutional Review Board. Participants were paid for their participation or completed the study on a volunteer basis. Participants completed a series of questionnaires aimed at characterizing their musical background. Specifically, participants completed the Barcelona Music Reward Questionnaire (BMRQ), which measures sensitivity to music reward. The BMRQ was used to ensure that no participants had specific musical anhedonia (BMRQ Score < 63) [38]. Note that people with musical anhedonia show decreased activity within the reward system in response to music, and this would likely affect the user's perception of the application [39]. None of the participants recruited had specific musical anhedonia, with the average sensitivity to musical reward of the participants (BMRQ mean plus sd: 86.9 ± 8.2) being slightly higher than that of the general population which is usually around 80 [38, 40]. Participants also completed the Goldsmith Music Sophistication Index - Musical Training subscale (Gold-

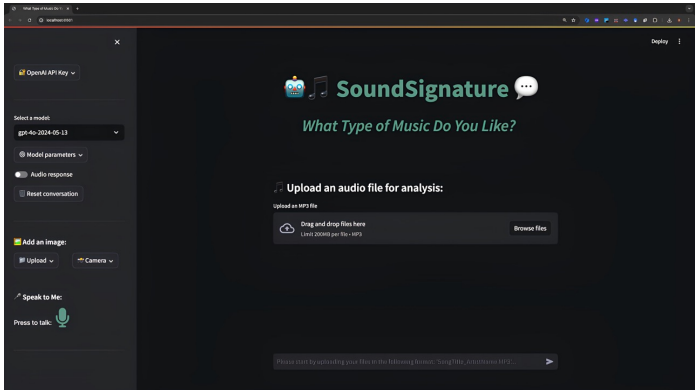

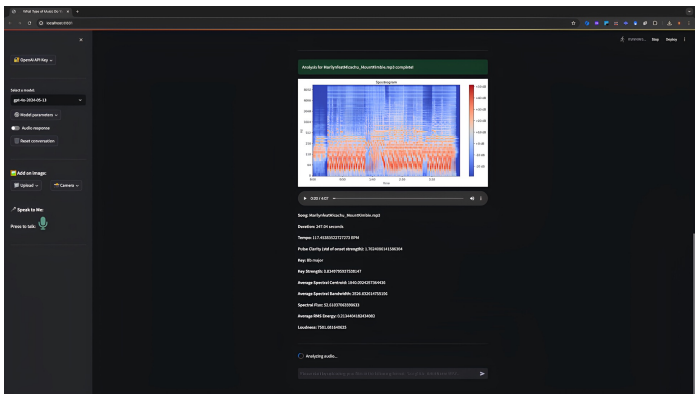

**Figure 1.** Screenshots from the *SoundSignature* web application during use

MSI-MT) [41], which quantifies the participants' amount of musical training received. The average musical training score was (mean plus sd) 28.6 ± 11.1, with this mean falling into the 53rd percentile of the general population. Participants' favorite music genres were diverse, although 55% placed rock, 35% placed jazz, and 25% placed pop and alternative among their top 3 favorite genres.

*B. Experimental Design*

Participants were first presented with an online consent form. They were then provided with a short overview of the procedures. Once online written consent was obtained, participants provided demographic information and then completed the BMRQ and the Musical Training subscale of the Gold-MSI. Participants then provided us with the names of 5-10 of their favorite songs (number of Songs mean plus std: 7.4 ± 1.9), which were obtained and formatted appropriately (i.e., SongName_ArtistName.mp3). We then walked the participants through the web application interface and uploaded their selected songs. Finally, once the analysis was completed and the assistant had responded, the participants were told to read through the response and let us know if they had any questions. Participants were also encouraged to interact with the chatbot further, and were provided with sample questions to ask (e.g., "What do these song choices say about me as a person?", "Based on these song choices, which other songs would you recommend?", etc.). Once they were done asking questions, they were required to complete a Likert scale from 1 (Completely Disagree) to 7 (Completely Agree) to rate how much they agreed with the Conclusion section provided by the app regarding their overall musical preferences (see Figure 1E). Participants were also asked to provide a Likert scale (also between 1 and 7) rating how much they agreed with the following statement: "This application informed me of my music taste in a meaningful way." Finally, we asked participants for their general thoughts on the app, allowing them to write as much as they wanted in an open-ended way. In addition to these written responses, the experimenter also took notes of other comments the participants provided during their interactions with the app. The survey containing the demographic questions, BMRQ and Gold-MSI questionnaires, and the questions about *SoundSignature* were all created and presented with Qualtrics Version 06/2024 [42].

IV. RESULTS

*A. Quantitative Responses*

The user study conducted to evaluate *SoundSignature's* efficacy and usability illustrated that participants were significantly satisfied with the application. Specifically, the average Likert rating for the agreement with the Conclusion provided by the app regarding the user's musical preferences was 6.3 ± 0.9. In addition, participants also felt that the app informed them about their musical taste in a meaningful way, with the average agreement score for this statement being 6.4 ± 0.7 (see Table 1).

*B. Qualitative Responses*

Responses to the open ended question were mostly positive (see Supplementary Table 1 for all responses). For example, participants positive responses included: "...it is pretty accurate and provided an insightful view of my musical tastes that I have not thought about before", "...felt very personal and told me a lot about myself and music preference", "...enjoyed reading the insights about myself based on my musical choices", "...I like how much more specific it gets once you ask follow up questions", among others. A theme that arose in the open-ended questions was the desire of users to know more about the relationship between the acoustic and musical characteristics of their favorite music and their mood and/or personality traits: "...what is the relationship between the kind of music that I listen to and my personality traits, mood, and feelings based on the tempo, spectral features, and lyrics…", "...could the app analyze the songs I have heard during a day and then tell me which kind of music and/or which kind of emotions I was looking for on that day?", "...what it could tell me in other areas of my life, such as movie taste, personality type…", "...tell me a bit more personal information, almost like a musical horoscope". One participant provided negative feedback, stating that "...most of the conclusions were correct and it informed me of my music taste but I think it doesn't work that well with Classical pieces…", while another participant mentioned that "...the first output is a bit generic".

TABLE 1

*Descriptive statistics describing how much participants agreed with the Conclusion section provided by the app regarding the user's overall musical preferences response, and how much they agreed with the statement "This application informed me of my music taste in a meaningful way."*

|  | Valid | Mean | Std. Deviation | Minimum | Maximum |
|---|---|---|---|---|---|
| Overall Musical Preferences | 20 | 6.300 | 0.979 | 4.000 | 7.000 |
| Informed Meaningfully | 20 | 6.450 | 0.759 | 5.000 | 7.000 |

In addition to the written open-ended responses, participants also shared their thoughts while interacting with the experimenter. Among these comments, we would like to highlight the following: "Did you feed my survey responses to the assistant before we started this? It's as if it's referring to questions I answered in the same way that I answered them", "When I asked for some recommendations based off of the songs that I uploaded, the first song it recommended was one I almost chose for my favorites, and it isn't a very well-known song", "I'm older, so many of the songs I chose are from the 70s and 80s. When I first asked it for recommendations based on my song choices, it recommended some of my other favorite songs. However, what was truly amazing was that when I asked it for recommendations from modern day artists, it gave a selection of songs from some of my son's favorite artists. It was really amazing to see the influence my listening habits had on him, and how we can both appreciate the modern day renditions of the music of my generation."

V. DISCUSSION

The *SoundSignature* application we present advances the state-of-the-art in the IoS by bridging MIR and semantic audio with AI and NLP, using cloud computing to provide users with personalized insights into their musical preferences in an innovative way. The results of the pilot user study suggest that the application was effective and highlights its educational potential and novelty within the MIR landscape. The results also underscore the application's ability to deliver meaningful and personalized insights into users' musical preferences. Participants agreed with the conclusion provided by the app regarding their favorite music and also felt that the app informed them about their musical taste in a meaningful way (in both cases the average rating agreement was above 6 in a scale whereby 7 means "Completely Agree"). Results thus suggest that the application effectively bridged the gap between complex music analysis and user-friendly interaction. Furthermore, several participants were excited about questions they could ask following the original output (e.g., "What do you think that the songs I chose say about me as a person?", "What careers do you think I might excel in based on my musical taste and the characteristics of my favorite songs?"; See Supplementary Materials for examples).

We hypothesize that the lyrical analysis and cultural and historical context provided by the Open AI assistant provided users with a holistic understanding of their musical taste. Instead of simply being told that they appreciate bright, high-energy songs in major keys, the assistant can provide reports such as "Your choices reflect a blend of introspection, social awareness, and cultural pride, suggesting that you value both personal and collective narratives in music." This approach aligns with contemporary research emphasizing the importance of context and personalization in music recommendation systems [21, 43]. By contextualizing the lyrical themes and the overlapping characteristics of the songs, *SoundSignature* might be able to offer a more personal and engaging user experience.

*SoundSignature* aims to instill a sense of confidence and control in the users, allowing them to manipulate and understand their favorite songs by simply interacting with the chatbot, echoing the findings of prior research which emphasizes the importance of interactive learning tools in music education . The integration of these features makes *SoundSignature* not only a tool for entertainment and personal use but also a powerful educational resource. Indeed, we believe that the educational impact of *SoundSignature* is one of its most noteworthy potential contributions. By translating complex musical data into understandable insights, the application aligns with educational theories that advocate using technology to make learning more accessible and engaging [44] and with the increased demand to include semantic audio features in IoS devices [22]. Users reported that the application provided them with information about their musical preferences and helped them understand the underlying musical concepts. A tool that provides insights into not only *what* type of music a person likes but also *why* they might like it has the potential to help clinical populations engaged in music-based interventions (MBIs) [45]. Indeed, MBIs can benefit from personalized and enriched music delivery systems where the stimuli used during therapy is individualized to the patient's preferences, thus potentially maximizing the effectiveness of the therapy provided [46].

This educational aspect will be essential in future versions of the app, as it can empower users with knowledge traditionally restricted to academic or professional settings, thereby fostering a more informed and engaged base of music listeners. This democratization of music analysis tools aligns with current pedagogical approaches where music serves as a contextual tool to facilitate deeper learning [16]. We hope the app will challenge the boundaries of traditional MIR systems by placing its primary focus on the user and by combining technical music analysis, educational content, and user-centric interfaces to cater to a diverse audience ranging from casual listeners to music enthusiasts and professionals [20].

We believe that *SoundSignature's* capability extends beyond personal use: it might also serve as a tool for building extensive music databases. The application can facilitate the creation of a rich music database by storing the musical and acoustic features alongside the metadata about the artists and lyrics it extracts. A scaled version of the app that allows users to create accounts and create a personal database could support more nuanced music recommendations that are based on acoustical and musical properties, but also on mood, themes, and even lyrical content, going beyond traditional recommendation algorithms. This capability could be especially beneficial if users started giving the chatbot more feedback in their responses or if the app offered "thumbs-up" and "thumbs-down" responses. Then, users could tell the chatbot which recommendations they like, whether something was inaccurate in the output, etc. Moreover, if we allow users to connect with their friends on the app, we could introduce shared reports that compare and contrast their *SoundSignature* profiles. This feature would foster more interactivity between users, enabling social engagement through music. For instance, users could explore shared song recommendations or create collaborative playlists based on their combined music profiles. This would align with the broader vision of the IoS and IoMusT, where interconnected users and devices collaborate to enhance creative exploration and personalization [22, 47]. By integrating social elements into the app, SoundSignature could leverage the IoS ecosystem to provide a more dynamic and interactive user experience, influencing how users engage with music and with each other.

Similarly, this application could label data for music datasets used for deep learning. Generative applications like *Riffusion* [48] and *MusicLM [49]* use diffusion models to create music from text inputs and are trained on spectrograms along with their labels. *SoundSignature* offers a unique opportunity to enrich these datasets by capitalizing on the feature set it extracts. By associating natural language descriptions and detailed musical and acoustic analyses with traditional spectrogram data, *SoundSignature* could provide richer, more contextual labels for training deep learning models for music generation. Being trained on a dataset with enriched labels could enhance a model's ability to generate music that resonates more deeply with human emotional and aesthetic preferences. Picture a model that could generate new compositions by interpolating and extrapolating from the features and styles of music the user prefers. For instance, users could upload their favorite songs, and *SoundSignature* could use the aggregated data to generate a new song. This song could mimic the input tracks' style, texture, and chords and attempt to incorporate the underlying emotional elements that the analysis has identified as significant to the user. Such a capability would be an exciting addition to the personalized music technology landscape, moving from passive analysis to active creation of user-specific content. This addition would also align with emerging trends in AI, where the focus is shifting towards more personalized, context-aware applications

[21, 43]. In addition, creating datasets that include enriched labels could also be useful for IoS devices and ecosystems, as they strongly depend on databases that include sound and music-related information to function [22]. The foundation laid by *SoundSignature's* data analysis capabilities provides a promising starting point for such developments.

*A. Limitations*

While the application has shown promising results, several limitations must be addressed. First, on closer inspection of the tempo analysis for the songs provided by participants in the pilot study, we realized that the tempo detection algorithm employed using *librosa* sometimes miscalculated BPM in half-time or double-time, which can misrepresent the song's actual tempo. This error can cause the chatbot to refer to a song as slow and laid-back when it might actually be upbeat and energetic, or vice versa. While there are other machine learning tools to estimate tempo (i.e., *madmom*) [50], these algorithms are computationally expensive and take longer to run, which can be a caveat and decrease the interactivity and ease-of-use of the app.

Additionally, the initial responses from the chatbot can sometimes be generic, as mentioned by one participant in the open-ended responses. However, when users engage with the system by asking more detailed questions (e.g., "What do these song choices say about me as a person?) the responses become significantly more personalized and insightful (see the qualitative results section). This indicates that while the chatbot is capable of deeper analyses, its effectiveness in providing the user with deeper details or ideas it has formed is contingent on user interaction, suggesting a need for more intuitive query handling and response generation.

Another limitation is that the OpenAI model the custom assistant runs on is only trained on data until October 2023. While this still allows for millions of songs and thousands of bands to be accounted for, it also means that the assistant may have difficulties providing details for newer artists. Interestingly, though, when conducting initial experiments, we uploaded an unreleased song by an unknown artist, and the chatbot was still able to provide interesting insights about the song and could somehow infer which genre it was (progressive rock) based on the features it was fed. Thus, the sections of the output that this knowledge cutoff date affects are the Cultural and Historical Context and Lyrical Content sections.

A notable limitation of our study is the predominance of participants with a Western musical enculturation and the selection of primarily Western songs. This bias may limit the generalizability of our findings, as musical preferences and perceptions can vary significantly across different cultures and musical traditions. While it did still seem to accurately report on the non-Western songs that were chosen, the insights provided by *SoundSignature* may not fully capture the diversity of musical tastes and preferences present in non-Western contexts. Future studies should aim to include a broader range of musical styles and participants from diverse cultural backgrounds to enhance the applicability and inclusivity of the application's analyses and recommendations. Finally, future user studies should include a control condition to ensure that the conclusions provided by the app are actually specific to the participant (e.g., provide participants with two app generated summaries of their musical taste, one stemming from the songs they selected and another from songs selected by another participant) [51, 52].

VI. FUTURE DIRECTIONS

Looking forward, several enhancements are planned for *SoundSignature*. We chose a number of features to be extracted capitalizing on previous MIR, music psychology, and music neuroscience literature [30-32]. Nonetheless, there are many other available acoustic and music features that can be extracted (e.g., Attack Time, Roughness, Fluctuation Centroid and Entropy, etc.) and added to the pipeline. While we can infer from the user study that users are satisfied with the current state of the application, we plan to carry out further A/B testing to see whether people prefer a different combination of features or different iterations of the prompt that we provide the OpenAI Assistant with. Incorporating additional features may or may not enrich the data fed to the chatbot, and slight alterations to the prompt might significantly change how the assistant responds.

We also plan to conduct user studies with artists, composers, and producers who could provide insights into the app's utility for professional users with advanced musical knowledge, helping refine the tool's accuracy and user interface. This would also allow us to understand whether the application is meeting the current demands of the music industry. A music producer might utilize the source separator to sample the drums or vocals for a new beat. In contrast, a classical composer may first utilize the source separator to extract the strings section of a piece and then run the audio-to-midi on that stem to notate the piece and conduct a harmonic analysis. We want to ensure that advanced questions and function calls are accurate, run smoothly, and meet the needs of potential users.

Eventually, we would also like to add measures related to cognitive processing to *SoundSignature's* analysis pipeline, such as musical surprisal. Over the past several years, research on musical complexity and surprise in cognitive neuroscience and music psychology has gained traction, leading to computational models that calculate prediction errors from music (e.g., IDyOM) [53]. The concept of musical surprise ties into theories proposing that an inverted U-shaped relationship exists between complexity and listener preference [54, 55]. These theories suggest that there is an optimal level of complexity that is variable for each individual depending on their musical experiences, musical training, etc. This variability implies that each listener has a unique internal model of what distinguishes something as complex, where one person's sweet spot might be boring or overwhelming to another. Thus, a future version of the app could include a measure of musical complexity. For example, we plan to include the Dynamic Regularity Extraction model (D-REX) to the app's pipeline [56]. D-REX is a Bayesian model of auditory salience that processes auditory features such as energy, pitch, and temporal modulation of continuous sounds to calculate surprisal scores, factoring in cognitive constraints like finite working memory and observation noise [56]. While IDyOM is similar in its approach to modeling musical surprisal, it only works with MIDI data, and thus cannot be applied to full songs [53]. Furthermore, recent findings from our lab show that the D-REX output correlates with both behavioral and neural measures of musical surprise

[57, 58]. In other words, the model ratings of surprisal can be used as a proxy for musical complexity. The addition of the D-REX measures would add quite an interesting dimension to the report (e.g., "Your music choices reflect a preference for music that is unpredictable and complex.").

Moreover, integrating personality assessments could provide fascinating insights into the relationships between musical preferences, personality traits, and acoustic features, broadening the scope of personalized music recommendations and psychological studies in music perception [59-65]. Note that several participants showed an interest in getting information about how the output of *SoundSignature* relates to their personality traits and/or mood state (see qualitative results section). Much music preference research has examined the relationship between personality and music preference. However, many findings are incongruent and merely focus on associating broad personality traits with preferences for specific genres [59]. Interestingly, research has consistently demonstrated that specific musical features elicit distinct patterns of neural responses (fMRI) in participants with a similar musical enculturation. For example, timbral features have been associated with activations in cognitive areas of the cerebellum and sensory networks, while the processing of musical pulse and tonality recruits both cortical and subcortical circuits involved in cognitive, motor, and emotion-related functions [32, 66]. Such findings suggest there may be a neural basis for the interaction between acoustic features and emotional and cognitive processing during music listening, at least within participants with the same musical enculturation.

## VII. CONCLUSION

The results from the user study suggest that *SoundSignature* effectively fulfills its objectives of providing deep, personalized insights, and showcases the app's potential to serve as an educational platform. These findings lay a promising foundation for future enhancements and the wider application of the technology. By addressing its current limitations and exploring new avenues for user engagement, *SoundSignature* is well-positioned to become a key tool for music lovers and professionals alike, reshaping the landscape of music analysis and appreciation for the general public including clinical populations [45].

As part of the broader IoS and IoMusT ecosystems, *SoundSignature* contributes to the development of interconnected musical tools and platforms that allow users to engage with music in new, interactive ways. By leveraging cloud-based processing and offering intuitive access to advanced music analysis tools without requiring technical expertise, the app aligns with the IoS vision of a networked, interactive audio landscape that includes semantic audio features. Through future expansions, such as integrating real-time interactivity and shared user reports, *SoundSignature* can further enhance social connectivity and creative exploration through music, offering new possibilities for personalized, context-aware music experiences and enrich current music and sound-related databases . As a result, it not only advances the state of music technology but also contributes to the evolving IoS framework, making sophisticated music analysis tools accessible to a wider audience.

ACKNOWLEDGMENTS

Research reported in this publication was supported by the NCCIH of the National Institutes of Health under award number R34AT012943.